\documentclass[aps, superscriptaddress, twocolumn,floatfix,showpacs]{revtex4}

\usepackage{graphicx}
\usepackage{dcolumn}
\usepackage{bm}
\usepackage{amssymb}
\usepackage{siunitx}
\usepackage{natbib}

\begin{document}

\title{The Newtonian mechanics of a Vibrot}

\author{H.~Torres\footnote{harol.torres@fau.de}}
\affiliation{Institute for Multiscale Simulation, Friedrich-Alexander Universit\"at, 91054, Erlangen, Germany}
\affiliation{``Henri Poincar\`{e}" Group of Complex Systems, Physics Faculty, University of Havana, 10400 Havana, Cuba}

\author{V. M. Freixas}
\affiliation{Departamento de Ciencia y Tecnología, Universidad Nacional de Quilmes, Argentina.}

\author{D. P\'erez}
\affiliation{Instituto Superior de Tecnolog\'ias y Ciencias Aplicadas, Cuba.}

\date{ \today}

\begin{abstract}

A mechanical model was developed to describe the behaviour of a device able to transform vibrations into rotations, named Vibrot. The theoretical model, developed in the newtonian formulation of mechanics, was able to reproduce qualitatively all the experimental results existing in the literature, and quantitatively some of them.
\end{abstract}

\maketitle

\section{Introduction}

Devices able to transform rotational energy into kinetic energy along the rotation axis have been very used for ages. Archimedes of Syracuse is credited with the invention of one of these devices, nowadays known as ``Archimedes's screw'', whose basic working principle is still used \cite{Screw0,Screw1,Screw2}. This kind of phenomenon have been well studied because of its direct implications in the industry, but what about the inverse phenomenon? The mysterious rotation of the statue of Neb-Senu, \cite{Statue}, attracted the attention of several believers who explained it by formulating fanciful hypothesis. Nevertheless, the physicist Brian Cox formulated a more realistic one, where the vibrations produced by the steps of the visitors \emph{induced} a rotation on the statue, \cite{Statue}.

In order to unravel the working principle behind this phenomenon we will focus our attention in a simple device named Vibrot (\emph{vib}ration to \emph{rot}ation). A complete experimental description of a Vibrot as the one we aim to describe theoretically, may be found in \cite{Altshuler}.

\section{The model}

\begin{figure}
  \centering
  \includegraphics[width=0.48\textwidth]{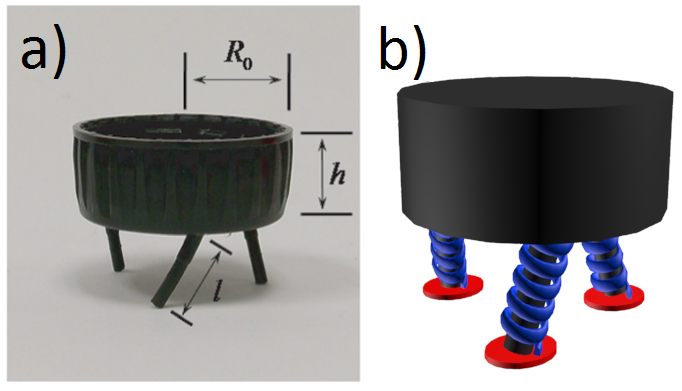}\\
  \caption{a) Photograph of a Vibrot taken from \cite{Altshuler}, and b) Sketch of our model Vibrot.}\label{vibrots}
\end{figure}

A Vibrot as the one designed by Altshuler \emph{et al.} in \cite{Altshuler}, is shown in Fig. \ref{vibrots}a). We model a Vibrot as a head resting on three legs. The head consists in an homogeneous cylinder of radius $R=15$ mm and height $h=12$ mm. Each of the legs is made by a spring of equilibrium length $l=11$ mm and elastic constant $=400$ N/m. The upper side of the springs are attached to the Vibrot's body at a distance $r=12$ mm from the center of the cylinder's bottom. A light rod freely slides inside each spring, and touches the ground through a flat disk. The three legs are axially symmetric relative to the bottom base, and are inclined $\alpha=30^\circ$ relative to the vertical  (see Fig. \ref{vibrots}b)).

The vibrations of the platform used in \cite{Altshuler}, can be introduced by means of an acrylic surface that vibrates sinusoidally in the vertical direction (i.e. along the gravity), with a frequency $f_v$ and amplitude $A$. Our reference system was taken on the vibrating surface, so it is an accelerated framework. In order to describe the temporal evolution of the Vibrot we use the following coordinates:

\begin{itemize}
\item $z(t)$: axis perpendicular to the base of the cylinder passing through its center.
\item $\varphi(t)$: rotation angle of any point located in the cylinder's base.
\end{itemize}

We first describe the forces acting along the  z-axis. The gravity force $F_g=mg$ points downwards, where $m$ is the Vibrot's mass, and $g=9.8$ m/s$^2$ is the acceleration of the gravity. When the springs are compressed a length $\Delta l$, an elastic force $F_e=-3k\Delta l \cos(\alpha)$ is applied upwards. That happens when the distance between the base of the cylinder and the ground is smaller than $l\cos(\alpha)$. Otherwise the Vibrot is in its ``flying phase'' and $F_e=0$, while springs keep their equilibrium length. As a consequence of taking an accelerated reference system we also have an inertial force, $F_i=m\omega^2A\cos(\omega t)=mg\Gamma\cos(\omega t)$, where $\Gamma=A\omega^2/g$ is the dimensionless acceleration and $\omega=2\pi f_v$ the angular velocity.

It is important to take into account, that every time the Vibrot collides with the platform, part of its kinetic energy is dissipated as heat and, like there is not previous models of a Vibrot, we will assume it happens due to a viscous friction. Strictly, this viscous force regards also the interaction between the Vibrot and the surrounding air, but this contribution can be neglected (the change in the kinetic energy by \emph{``air-friction''} is much less than by \emph{``platform-friction''}), so we assume that the viscous friction only mimics the collision between the Vibrot and the acrylic surface. Then, in the z-axis there is also a viscous force $F_v=-b\dot{z}(t)$, where the parameter $b$ will be determined from the experimental data.

If we define the variable $\xi(t)=z(t)-l\cos(\alpha)$, the viscous and elastic forces can be written as

\begin{equation}\label{Fv}
  F_v = -b\dot{\xi}(t)
\end{equation}

\begin{equation}\label{Fe}
F_e = -k\xi(t)
\end{equation}

As we mentioned before, the elastic force only acts when the Vibrot is touching the platform ($\xi<0$). To introduce this information we can use the Heaviside function, $\Theta(x)$. Then, the elastic force is $F_e=-k\xi(t)\Theta(-\xi)$. Henceforth we will discuss the motion in the z-axis by means of the variable $\xi(t)$. We will also define $\omega_0=(2\pi f_r)=\sqrt{3k/m}$, where $f_r$ is the rotation frequency.

So, the differential equation describing the vertical motion of the Vibrot is

\begin{equation}\label{eqofmot}
  \ddot{\xi}(t)=g\Gamma\cos(\omega t)-g-\omega_0^2\xi(t)\Theta(-\xi)-\frac{1}{m}b \dot{\xi}(t)
\end{equation}

In order to obtain the equation that describes the rotation of the Vibrot, we will pay attention to the momentum of the forces exerted on the cylinder. When the elastic force is acting, it has a component on the plane of the base of the cylinder, which has associated an absolute torque, $T_e$, given by

\begin{equation}\label{elastictorque}
  T_e=3k|\Delta l|\sin(\alpha)r=3k|\xi(t)|\tan(\alpha)r
\end{equation}

Both the torque and the elastic force are only different from zero when $\xi(t)<0$. Furthermore, once the Vibrot touches the platform, a kinetic friction force begins to act, but quickly transforms itself into static. Modelling this force is also complicated, but we propose the following hypothesis: \emph{once the Vibrot collides with the vibrating membrane, the torque mentioned before changes its sign, and remains this way until it reaches the minimum, where it comes back to be positive}. Moreover, as the static force cannot move the device backwards, we will assume the angular velocity to be non-negative. Also, we include a dynamic friction associated to the torque, $T_{Fr}$, given by Eq. (\ref{frictiontorque}), where was taken as average normal force the ones in the position of equilibrium of the springs.

\begin{equation}\label{frictiontorque}
T_{Fr}=-\mu m g r
\end{equation}

The friction coefficient was taken as, $\mu=0,5$, which is the very close to the one between acrylic and caoutchouc. In order to avoid that the legs of the Vibrot slide backwards, the relation $\mu_{static}>\tan(\alpha)$, must be fulfilled all the time. Note that if the friction coefficient is null, wherewith the Vibrot couldn't rotate and due to the external torque vanishes while the angular momentum conserves, then the rotation frequency remains with the initial value, \emph{i.e.} $f_{r}=0$.

The differential equation that describes the temporal evolution of the rotation angle, $\varphi(t)$ is

\begin{eqnarray}\label{eqofmotfi}
  \nonumber \ddot{\varphi}(t) & =-\frac{2r}{R^2}\omega_0^2\dot{\xi}(t)\tan(\alpha)\Theta(-\xi(t))\Theta(\dot{\varphi}(t))Sig(\dot{\varphi}(t))-\\
  & -\frac{2r}{R^2}\mu g \Theta(-\xi(t))\Theta(\dot{\varphi}(t))
\end{eqnarray}

In this equation, $Sig(x)$ represent the sign function, and we used as the moment of inertia of the Vibrot, $I=1/2mR^2$ (Moment of inertia of a cylinder). Equations (\ref{eqofmot}) and (\ref{eqofmotfi}) describe the temporal evolution of the device. We will solve them assuming that the Vibrot is at rest when $t=0$, so the initial conditions are:

\begin{equation}\label{initcond}
   \dot{\varphi}(0)=0\qquad \xi(0)=-\frac{g}{\omega^2}\qquad \dot{\varphi}(0)=0\qquad \varphi(0)=0
\end{equation}

\section{Results}

Solving analytically the system of differential equations (\ref{eqofmot}) and (\ref{eqofmotfi}) is a little complicated in spite of having the first equation disengaged. So, we have to solve it numerically. However, we can reach some analytical results, as follows.

In our model, the Vibrot does not rotate if the legs are touching the ground. So, if $\xi(t)$ is negative, the Vibrot will not rotate, and it is possible to determine the threshold dimensionless acceleration $\Gamma_{th}$, below which there is no rotation. In order to do that, the equation (\ref{eqofmot}) is solved for $\xi(t)<0$, resulting the equation of a forced oscillator plus a constant. Then if the solution for a long time ($t\gg\frac{1}{f_{v}}$) is negative, the Vibrot will not rotate. For long times, the solution that prevails is:

\begin{equation}\label{etapart}
  \xi_{p}(t)=\frac{g\Gamma \sin(\omega t+\vartheta)}{\sqrt{(\omega_0^2-\omega^2)^2+\left(\frac{b}{m}\right)^2\omega^2}}-\frac{g}{\omega_0^2}
\end{equation}

where

\begin{equation}
  \vartheta=\arctan\left(\frac{\omega^2_0-\omega^2}{\frac{b}{m}\omega}\right)
\end{equation}

By imposing $\xi(t)=0$  and $\sin(\omega t+\vartheta)=1$ in Eq. (\ref{etapart}), it is possible to obtain $\Gamma_{th}$ as:

\begin{equation}\label{gammath}
  \Gamma_{th}=\frac{\sqrt{(\omega_0^2-\omega^2)^2+\left(\frac{b}{m}\right)^2\omega^2}}{\omega_0^2}
\end{equation}

\begin{figure}
  \centering
  \includegraphics[width=0.41\textwidth]{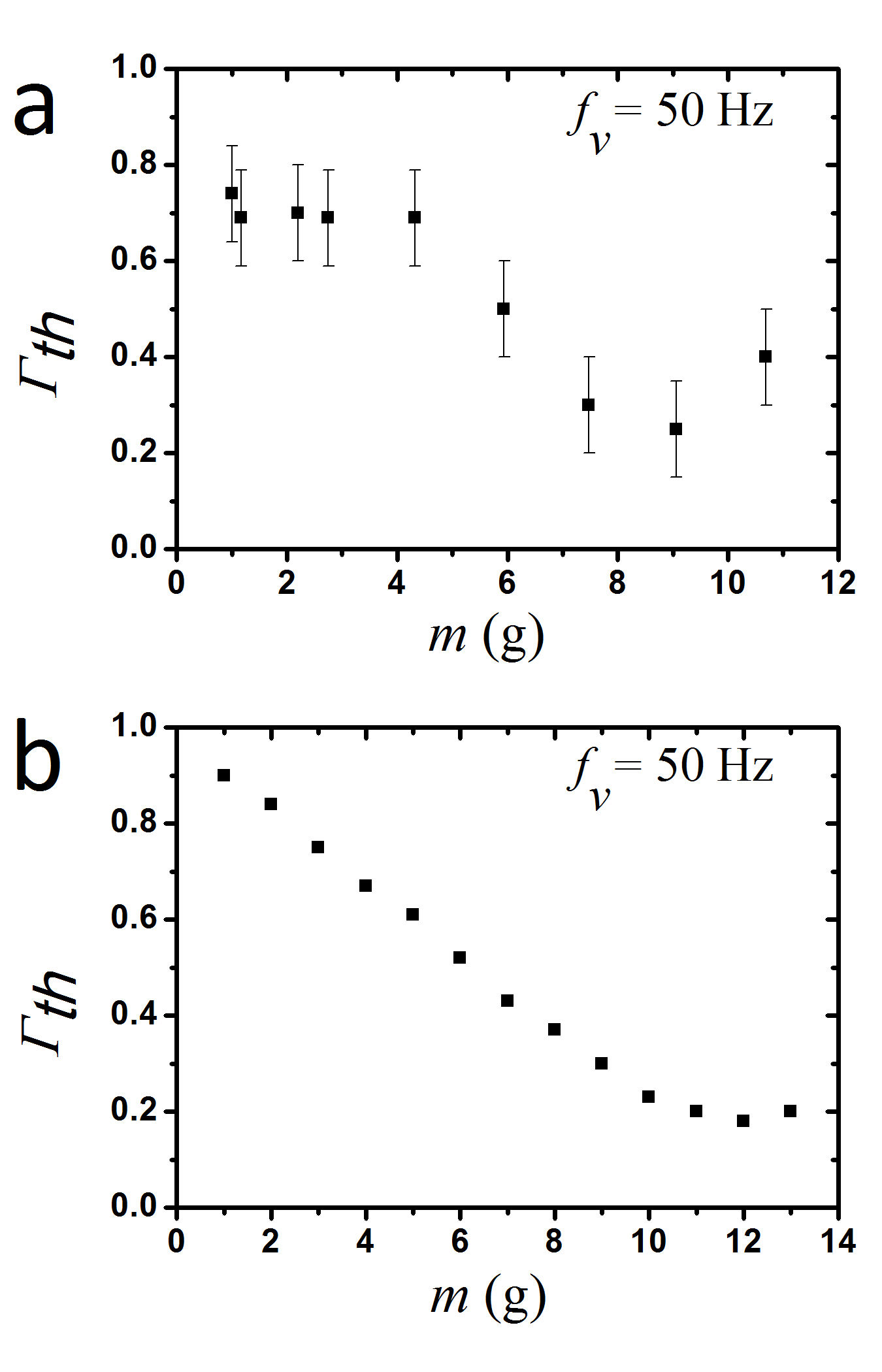}
  \caption{Comparison of the dependency of the threshold dimensionless acceleration, $\Gamma_{th}$, as a function of the Vibrot's mass, $m$, for a $50$ Hz vibration frequency, between a) Experimental results \cite{Altshuler} and b) Theoretical results.}\label{graphgammath}
\end{figure}

Using in Eq. (\ref{gammath}) the expression $\omega_0=\sqrt{3k/m}$, we obtain the dependency with the mass of $\Gamma_{th}$. Fig. \ref{graphgammath} shows the experimental result for $\Gamma_{th}$ vs. $m$, reported in \cite{Altshuler} as well as the one obtained here using Eq. (\ref{gammath}).

In Fig. \ref{graphgammath} we can see the resonance effect in both graphs. The model's results are slightly right-shifted in a value on the order of $3$g, a consequence of the choice of the parameters $k$ and $b$. Here, it is important to remark the almost perfect coincidence between the numerical results and the ones obtained using Eq. (\ref{gammath}).

\begin{figure}
  \centering
  \includegraphics[width=0.41\textwidth]{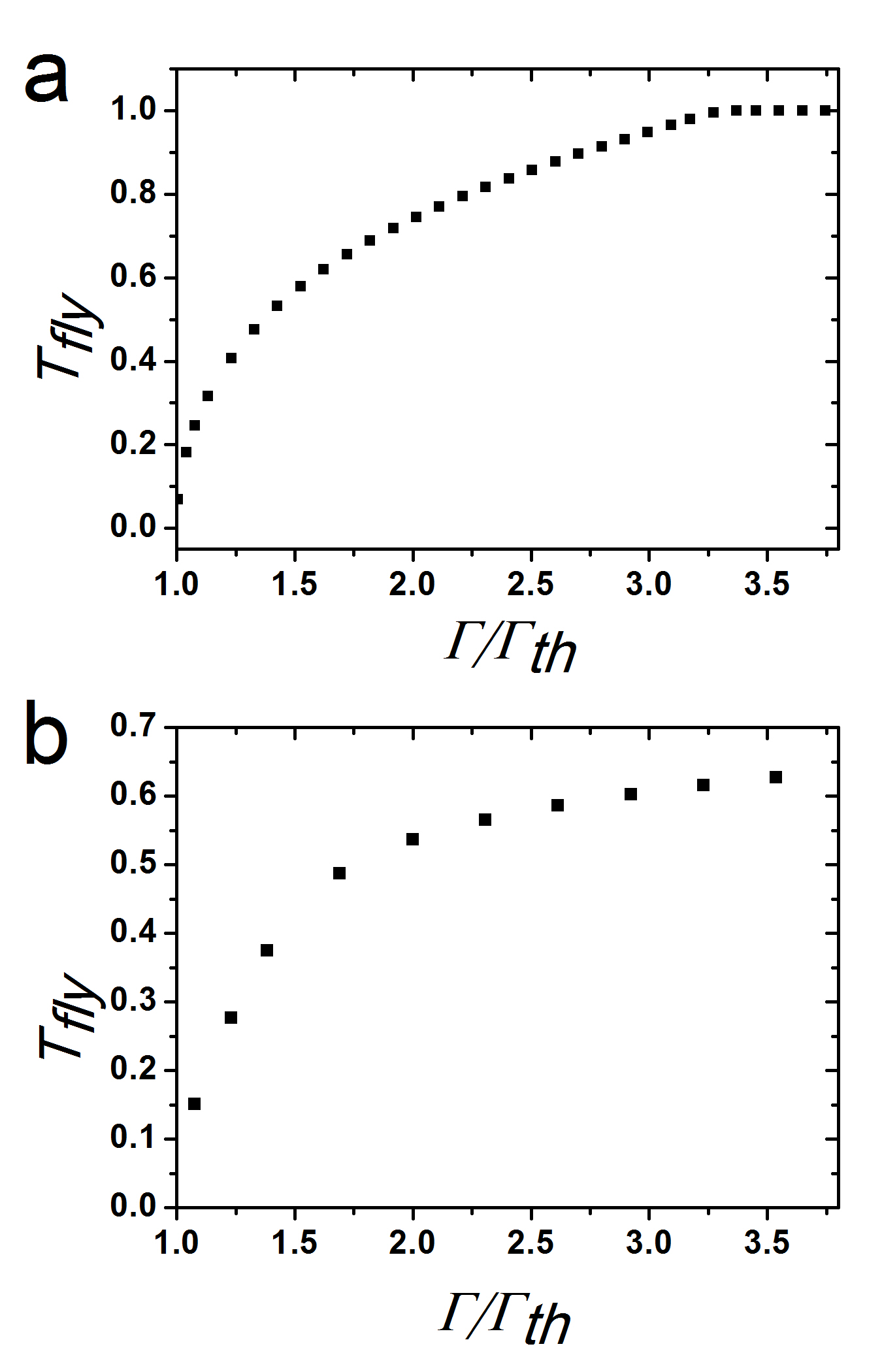}
  \caption{Comparison of the dependency of the fly time as a function of the dimensionless acceleration normalized to the threshold dimensionless acceleration, between a) Experimental results and b) Theoretical results.}\label{graphtfly}
\end{figure}

We also obtained an approximation for the fly time, $T_{fly}$, using the following expression:

\begin{equation}\label{tfly}
  T_{fly}=\frac{2\dot{\xi}(t_n)}{g}
\end{equation}

where $t_n$ is the n-th zero of $\xi(t)$, ($n\gg1$). We can see the fly time as a function of the normalized dimensionless acceleration ($\Gamma/\Gamma_{th}$) in Fig. \ref{graphtfly}. This figure shows that both graphs have the same behaviour, although we may say that the model underestimates the fly time.

\begin{figure}
  \centering
  \includegraphics[width=0.48\textwidth]{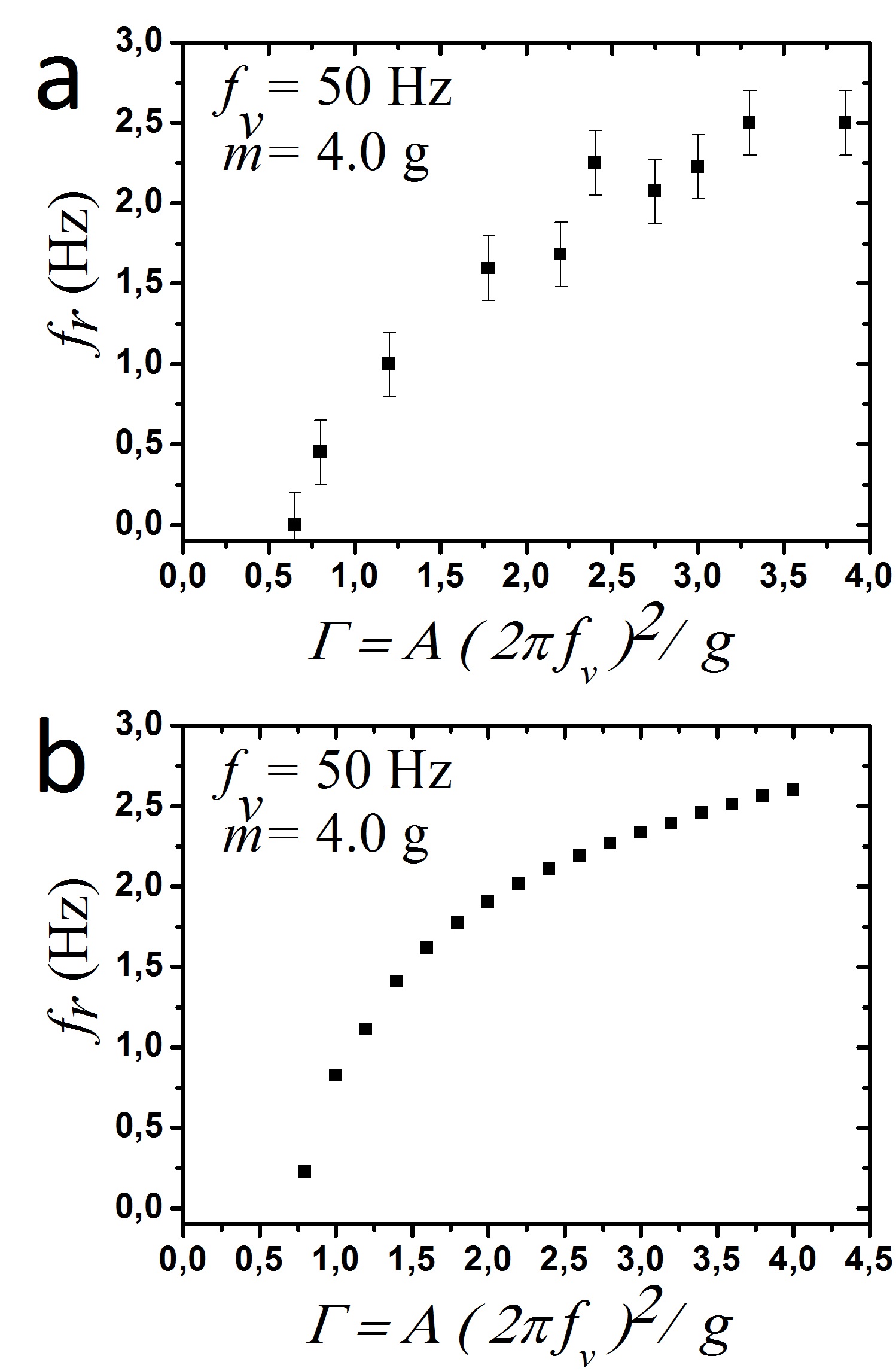}
  \caption{Comparison of the dependency of the rotation frequency as a function of the dimensionless acceleration, for a $4$-gram Vibrot under a $50$ Hz vibration frequency, between a) Experimental results and b) Theoretical results.}\label{graphfrvsgamma}
\end{figure}

Once solved numerically the system of differential equation, (\ref{eqofmot}), (\ref{eqofmotfi}) with boundary conditions given by (\ref{initcond}), it is possible to obtain the behaviour of further magnitudes, like the rotation frequency as a function of other parameters. Experimentally, the rotation frequency is computed as the average value of the function: $\dot{\varphi}(t)/(2\pi)$, but inasmuch as this function is not constant, we calculated the rotation frequency $f_r$ as:

\begin{equation}\label{rotfreq}
  f_r=\frac{1}{2\pi(t_2-t_1)}\int_{t_1}^{t_2}\dot{\varphi}(t)dt=\frac{\dot{\varphi}(t_2)-\dot{\varphi}(t_1)}{2\pi(t_2-t_1)}
\end{equation}

where the relations $t_1>0$, $t_2>0$ and $t_2-t_1\gg1/f_v$, have to be satisfied. Fig. \ref{graphfrvsgamma} shows a comparison between the experimental and the model results for the $\Gamma$ dependence of the rotation frequency, for a Vibrot with mass $m=4$ g and a vibration frequency of $50$ Hz. For the value of $\Gamma_{th}=0.7$ the model accurately predicts the experimental results.

\begin{figure}
  \centering
  \includegraphics[width=0.48\textwidth]{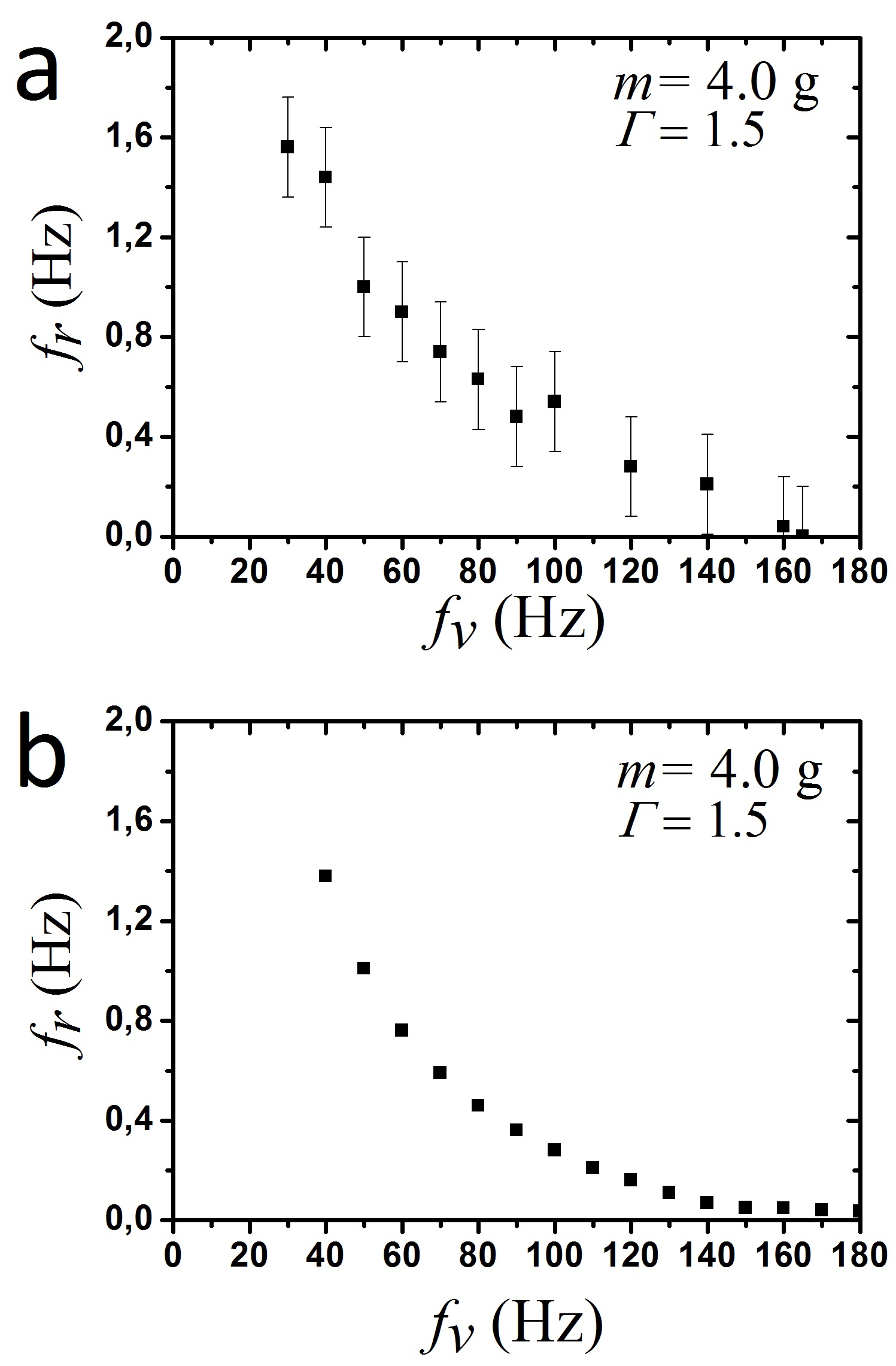}
  \caption{Comparison of the dependency of the rotation frequency as a function of the vibration frequency for a $4$ mass Vibrot under a dimensionless acceleration $\Gamma=1.5$, between a) Experimental results and b) Theoretical results.}\label{graphfrvsfv}
\end{figure}

In Fig. \ref{graphfrvsfv} we can see how the rotation frequency decreases as the vibration frequency increases, for a Vibrot with mass $m=4$ g and a dimensionless acceleration, $\Gamma=1.5$. Also, it shows a very good agreement between the experiments and the model. However, the latter predicts a ``smoother'' decay for high vibration frequencies.

\begin{figure*}
  \centering
  \includegraphics[width=0.9\textwidth]{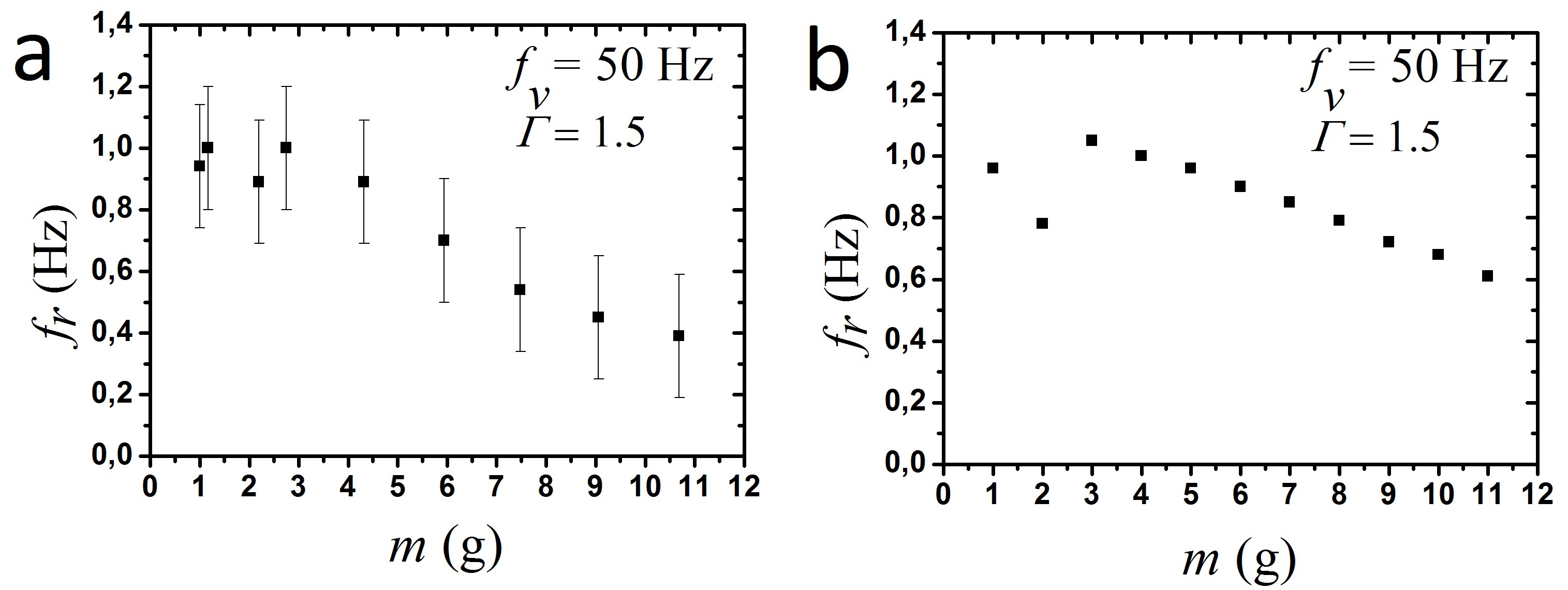}
  \vspace{-0.2cm}\caption{Comparison of the dependency of the rotation frequency as a function of the mass of the Vibrot, under a vibration frequency, $f_v=50$ Hz and dimensionless acceleration, $\Gamma=1.5$, between a) Experimental results and b) Theoretical results.}\label{graphfrvsm}
\end{figure*}

Our model also allows to obtain the mass-dependency of the rotation frequency, $f_r$. In Fig. \ref{graphfrvsm} we observe a very similar behaviour for small masses between the experiments and the model, but for higher masses the decrease of $f_r$ is slower for the model. This may be due to the fact that the rubber legs of real Vibrots are mechanically deformed for high masses.

It is important to check the model's results for the case of a Vibrot with rigid legs ($k\rightarrow\infty$). In this particular case, it was experimentally proven that the device does not rotate. This was an expected result since it behaves as a rigid body. With our model, we can corroborate this result, by giving big values to $k$ (bigger than $10^6$ N/m). As the elastic constant increases, the rotation frequency decreases, eventually reaching values very close to zero. This fact can be explained taking into account that an increasing of $k$ provokes an increasing in the resistance of the springs to the external force (vibrating platform), which implies a smaller amplitude, and bigger velocity and fly time, what finally results in a smaller rotation frequency.

\section{Conclusions}

We have developed a mechanical model of a Vibrot: a device that, when put on a vibrating platform, is able to rotate due to the presence of inclined elastic legs that interacts frictionally with the vibrating platform. In order to construct the system of differential equations of the model, the reference system was taken on the vibrating platform and the rubbery legs of the experimental Vibrot \cite{Altshuler} was modeled as linear springs inclined a certain angle relative to the normal of the base of the cylinder, which represents its body. The loss of energy mechanism was proposed as a viscous force in the vertical axis and a kinetic friction force in the angular direction. All the values to construct the model were taken from \cite{Altshuler}, except the coefficient of the viscous force, which was determined by fitting the experimental data and the friction coefficient, which was taken from the literature.

In spite of the relative simplicity of our model -entirely based in Newtonian mechanics- we have been able to reproduce semi-quantitatively or quantitatively most experimental observations on the Vibrots reported in \cite{Altshuler}. The few discrepancies with experimental data are only quantitative, an dare probably due to the fact that we did not model the mechanical deformations of the shape of the legs (especially bending) due to the weight of the Vibrot body.


\textbf{Acknowledgment}

We would like to thank to Diego Maza for providing experimental data.


\end{document}